\documentclass[twocolumn]{adonis}
\usepackage[hidelinks]{hyperref}
\usepackage{tabularx}
\usepackage{xurl}
\usepackage{graphicx}
\usepackage{listings}
\usepackage{xcolor}
\usepackage{graphicx}
\usepackage{makecell}

\title{Granularity in Action}
\subtitle{Graphing sources for social history}
\author{Sofus Landor Dam and Johan Heinsen \textsuperscript{1}}

\affiliation{\textsuperscript{1} MASSHINE, Aalborg University}
\correspondence{sld@society.aau.dk}
\version{\today}

\runningauthor{Sofus Landor Dam and Johan Heinsen}
\runningtitle{Granularity in Action}
\abstract{This working paper describes a pipeline for turning historical sources into structured data organised around the principle of foregrounding action as the basic and constitutive unit of analysis. It is rooted in a desire for pipelines that suit a granular approach to social history. The pipeline rests on the principles developed in the GRAM-framework (Graph of Roles and Actions Model), but leverages a range of machine learning tools to allow for an automated, skeletal graphing of actions. Ideally, such auto-GRAMS would integrate with close readings, including extensive manual graphing. Finally, we provide an example of how this approach might work in practice by graphing actions of pretending across four separate archival collections, relating to runaways and itinerants in eighteenth and nineteenth-century Denmark.

\vspace{0.3cm}

\noindent\textbf{\upshape Keywords:} \upshape Granular social history, graph database, archival reconstruction, machine learning
}

\begin{document}
	\maketitle
	
	\section{Introduction}
        
        As historians transform archival sources into research data, those data should have textures that align with the granularity of the materials from which they are derived. In this context, this means to work from the explicit descriptions of actions as embedded in verbs and their contexts. Preserving that texture might inform querying for a practice-oriented, \textit{granular} social history centered on what people did to one another, and the specifics under which they did it. That is not to deny the role of interpretation or theory, but to make a case for empirical studies to rest on a data structure that aligns with a conception of social relations as configurations of multiple distinct elements and practices that manipulate those elements \cite{worck}. In a way, we attempt to background theoretically informed (pre)suppositions, and instead let the actions of historical actors do the talking.
        
        This approach is in contrast to other recent methodological scholarship in digital history contexts, that instead foregrounds the role of theory when constructing data structures or computational tools for historical inquiry \cite{Piper_2026,Craig2024Designing}. These studies call for theory to be built-in and used as a guiding principle - especially when non-deterministic elements such as machine learning and artificial intelligence are involved in the workflow. We, instead, try to reduce the interpretive role of language models to a minimum and adjust the unit of analysis to its smallest and most meaningful unit - the action.
        
       Here, we will outline an approach designed to allow social historians to query textual representations of actions in ways that can incorporate substantial amounts of data (more than human eyes could otherwise process) without losing sight of granularity. This is not an argument for a quantitative history, but rather one that has the tools to look past the artificial constructs of quantitative/qualitative (or macro/micro) by partially aligning data sources in the same data structure, big or small. In so far as this is considered a question of scale, it only relates to the amounts of sources and not the ontology of what is studied \cite{de_vito_history_2019}.

        We developed this approach as part of our ongoing research project \textit{Run Away} in which we study what people did while on the run, typically from either labour relations or state institutions.\footnote{Funded by the Independent Research Fund Denmark and running from 2025 to 2027.} The project focuses on Denmark in the eighteenth and nineteenth-centuries and studies a range of people including convicts, soldiers, sailors, apprentices and servants. During the period we study, we see strong indicators that running away and becoming someone else went from being a very real possibility to being virtually impossible, but that this was a slow, gradual process owing to a host of factors. We study that process on the basis of situations that illustrate how people tried, succeeded or failed.
        
		The pipeline involves:
		
		\begin{enumerate}
        
			\item Turning historical documents into standalone texts. This includes OCR as well as a post-processing step that renders pages (such as a newspaper page with many news items) into separate texts.
			
			\item Classifying texts as a way of processing large archival collections and isolate texts of relevance. These are then merged into a new synthetic collection.
			
			\item Identifying descriptions of actions in the new synthetic collection. This involves identifying verbs and then, if necessary, filtering them.

            \item Turning those verbs into graph data leveraging generative models to identify subjects, objects and locations of actions.
            
		\end{enumerate}

        Each step has modular elements, meaning that it revolves around a form of processing that needs to happen before advancing, but not necessarily using the same models or wording for prompts. For this paper, we will not discuss model performance for each step at any length. Performance will inevitably vary for others according to the language and time of the sources as well as the specific capabilities of models and much adjustment is needed to port the pipeline to process sources from other historical contexts. However, we will describe some of the decisions we took to improve the output. No models tested performed perfectly, so readers should adjust their expectations. 
        
        In this context it is also worth noting, that we decided on a set of priorities early on in the development. One of these was that we would use local models and smaller architectures whenever possible. We were not able to do all steps without server-side language models, but steps 1, 2 and 3 above could also be done using LLMs. This, however, would add to the price of the workflow, making it harder to iterate and more expensive to reproduce. That cost would also be environmental and would in some cases entail using research funds to fund companies that openly challenge human value.
    
    \section{Harvest time: From image to text}
        Information on runaways is found scattered across a vast archival terrain. For our project we focused on four bodies of texts:
        \begin{enumerate}
			\item Newspapers containing, among many other things, runaway advertisements. These were digitised as part of the ENO-project (Enevældens Nyheder Online | News during Absolutism Online) by Johan and Camilla Bøgeskov with assistance from Sofus as well as our colleagues and students Anders Dyrborg Birkemose, Kamilla Matthiassen, Louise Karoline Sort and Louise Emilie Pedersen. This digitised corpus consists of more than 560.000 newspaper pages in the form of digital images derived from microfilmed copies. Using custom Transkribus models for both text and layout, these pages were turned into about 474 million words of text covering a range of newspapers from the period 1666 to 1849, during which period Denmark was an absolutist kingdom, though highly lopsided as most stem from the period after 1760. The digitised text is relatively high quality, with error rates that are typically less than 5 percent on word level. However, bad microfilm scans create lacunae, in which the OCR quality drops. The corpus and its creation has been described in further detail by Heinsen and Bøgeskov \cite{WorldInPrint}.
            \item By the mid-nineteenth century, runaway advertisements had largely disappeared from newspapers, as documented by the work of Anders Birkemose \cite{Efterlyst}. However, from 1867 onwards notices relating to runaways and unidentified itinerants were printed in Denmark's police gazette \textit{Politiets Efterretninger}. This print publication was digitised in 2025 by us in collaboration with Camilla Bøgeskov, Kamilla Matthiassen, Max Odbjerg Petersen, Anders Birkemose and a group of very enthusiastic students. We digitised every edition until 1890 from high quality scans provided by the Copenhagen City Archives, using the same pipeline as the ENO-corpus, but with even better results owing to uniform digital images. The result is about 16.000 pages. The dataset is known as the SPOR-dataset.\footnote{\url{https://hislab.quarto.pub/spor/}}
            \item Colonial newspapers carried runaway advertisements too. In a Danish context, the Caribbean colony of St Croix was covered by newspapers from 1770 onwards. A dataset has been created by students at Aalborg University collecting all runaway advertisements in surviving newspapers from 1770 to 1810. This dataset is published on the WORCK data platform and contains 685 unique advertisements \cite{St_Croix}. These texts are in English.
            \item Finally, we complemented these different, but highly systematic digitisation projects with data gathered over the last ten years by Johan Heinsen. This data was digitised manually in working various source collections and consists mainly of interrogations and sentences of runaway, typically prison breakers or military workers. This collection of manually curated court records contains 753 texts of relevance. They stem from the period 1720 to 1830. While the collection is haphazard in terms of how it came to be, and therefore cannot be argued to be full or representative of anything, it contains a lot of detailed information that might shed further light on what is found in the systematic collections.
            \end{enumerate}

        In addition to the transcription, we also decided on a further processing step. We wanted to proceed from segmented pages, that is standalone texts. For the ENO-project and the St Croix dataset, segmentation into texts had already been undertaken, while the SPOR dataset was processed using a combination of rule-based scripts and supervised machine learning models. The collection of manual transcriptions was already segmented into standalone texts but because of their wildly varying lengths, these were further split into 22.018 chunks of varying lengths using the Segment-Any-Text framework.\footnote{\url{https://github.com/segment-any-text/wtpsplit}} We will not detail these segmentation processes here, but note that the resulting data structure is one in which one observation is either one short text or, as is the case with the court records, a chunk of text representing something akin to a paragraph (for instance, the answer to a question), linked back to the original record. While not entirely homogeneous we found that some level of uniformity in terms of text length eased later work tasks.
        
        The transformation of such archival collections into digital text is necessary for all the following steps. However, in some cases, a researcher might not need to isolate the individual texts within the digitised mess of pages. For example, if working with a book of sentences, it would not necessarily be a problem that a page contains two verdicts lumped together if proceeding directly to the graphing of relations, and if all verdicts were of potential relevance to the research question. However, for us, this process was critical for the print collections especially, as they contained hundreds of thousands of pages combined, relating to everything that might be worthy of a news story, an advertisement or a police notice. Processing all of it would be too computationally taxing to be reasonable and would only move the problem of identifying the relevant bits within the harvested materials downstream in the workflow.

    \section{Threshing the archive}
        While the St Croix dataset had already been filtered in terms of relevance by dint of its provenance, the other collections contained many texts that were not about runaways. Isolating the relevant parts of these collections was critical. We approached this as a classification problem. In both cases we manually annotated samples of texts and used this labeled data as training material for a setfit-model \cite{setFit_artikel}. The setfit-architecture consists of two elements: 1. A sentence transformer which is fine-tuned to see similarities between texts with similar labels; and 2. a logistic regression model that uses the embeddings provided by the fine-tuned sentence transformer to predict if a text fits the label or not. We have found that for texts following somewhat strict conventions - such as runaway notices – this technique works very well. In many cases off the shelf multilingual models might be the basis of such a model, but for our material we used a custom sentence transformer based on Old News Bert – a BERT-style model trained on the ENO corpus \cite{ON_bert}.\footnote{The sentence transformer fine tune of Old News Bert is found at:\url{https://huggingface.co/JohanHeinsen/Old_News_Segmentation_SBERT_V0.1}} This model was judged to be an especially good fit, given the overlap in source material provenance. It also performed well in a benchmark study on different materials from the same period \cite{Lassche_et_al}. It seems to have provided a good basis; the resulting models had very high F1-scores suggesting that they both had a high success in retrieval (not missing too many texts) and precision (not producing too many false positives).\footnote{For newspapers: \url{https://huggingface.co/JohanHeinsen/ENO_Runaway_Advertisement_classifier_2.0}. For police notices: \url{https://huggingface.co/JohanHeinsen/PE_efterlyst_classifier_v2}}

        \begin{figure}

            {\centering \includegraphics[width=0.45\textwidth]{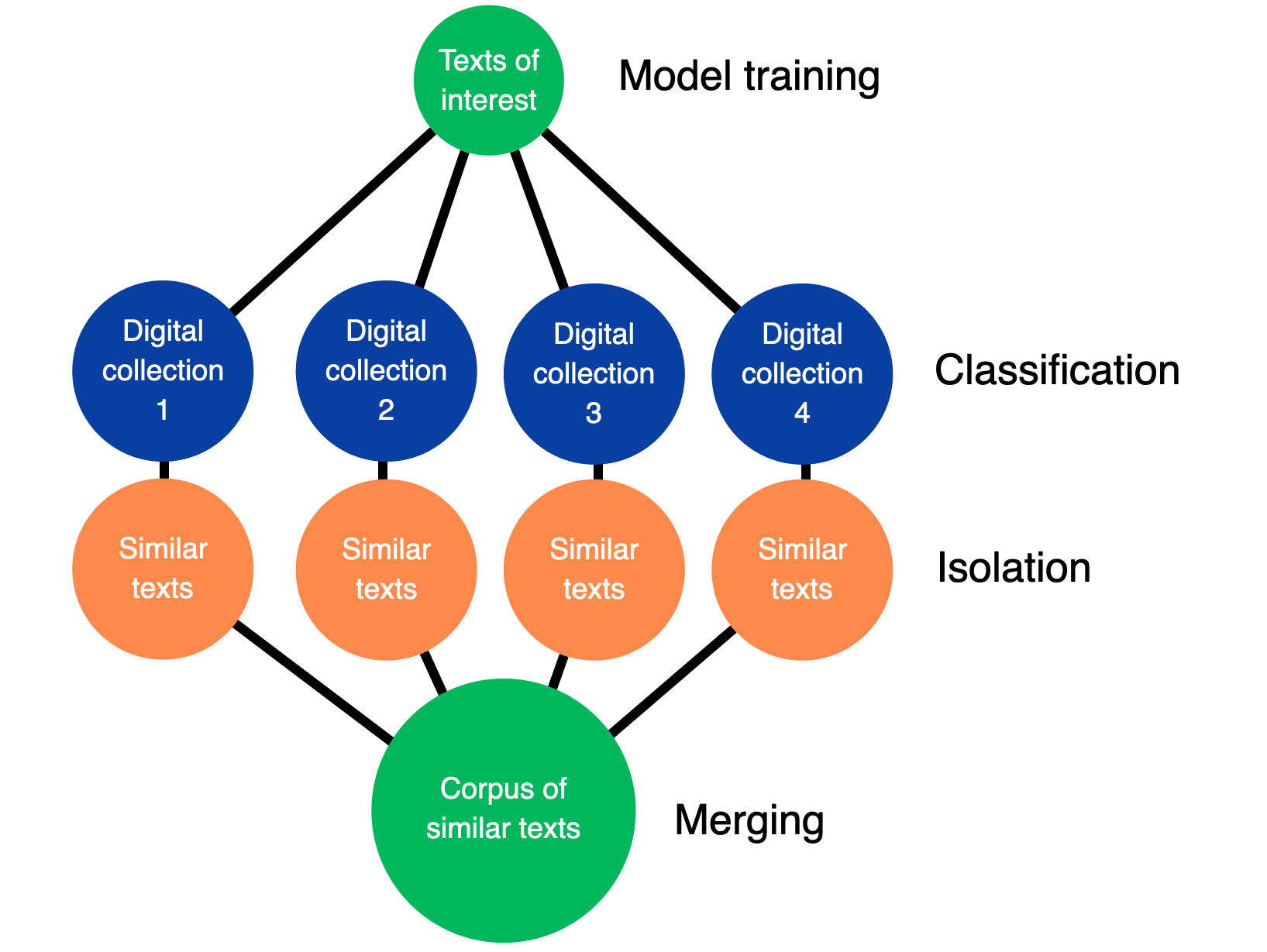}}

            \caption{Principle of archival threshing.}

        \end{figure}
        
        The texts identified by the classification models for ENO and SPOR were in the end combined with the St Croix texts and the manual transcriptions. ENO yielded 17.430 texts and SPOR 22.465. The two sets were not identical in terms of genre. The advertisements in the ENO-corpus almost all relate to runaways from employers (both state and private). Meanwhile the people advertised in the police notices of SPOR included people escaped from coercive labour (convicts, military deserters), but also many runaways in a broader sense, typically strangers whose identities were unverified, but who was now to be apprehended or whose identities needed untangling. Together with the text chunks from the manual transcriptions, they came to form the basis of a new collection to be graphed for our projects. 
        
        Fundamentally, this classification step is a crude operation. It does not rely on reading either along or against the grain, but in a blunt, persistent maneuver, fundamentally dependent on a tool. We think of it like a kind of \textit{threshing}, meant to separate some element from a whole, but knowing that for it to scale, some loss is inevitable.

    \section{Milling texts}
        The notion of breaking sources down according to verbs is heavily inspired by the 'verb-oriented' method developed by Maria Ågren and her team in the \textit{Gender and Work}-project at Uppsala University \cite{agren_making_2017}. However, we take a much more literal approach. This starts by using a part-of-speech tagger which classifies individual word tokens. For many languages such models are readily available, but we found that models built for modern Danish struggled with early modern Danish, and perhaps its idiosyncratic capitalization especially. For this reason, Sofus trained a tailored tagger called HERMOD (Historical Entity Recognition Model).\footnote{GitHub link:\\\url{https://github.com/CALDISS-AAU/BP_HERMOD}} Again, this model derives its features from the above-mentioned BERT-model and was based on hand-annotated examples of texts from the period. 
        
        We apply this model to the text pieces. It outputs the character spans of verbs in the texts. The output is then used to split the dataset so that each observation becomes an individual verb-token. The basic idea is that each verb is somehow related to an action and that the character spans can help guide later processing. That is obviously not always true, as some actions might take several verbs to describe.

        For the English language dataset from St Croix, we found that an off-the-shelf part-of-speech tagger from the SpaCy-framework worked well enough to process the material and isolate verb-instances. Such evaluation will have to be undertaken on a case by case basis.
        
        In our project, the texts contained more than 300.000 Danish verb-instances, many of which we judged were not of particular relevance to our project. This included auxiliary and modal verbs, but also a lot of verbs that typically related to the advertisers' requests to the public and not the running away. Since subsequent steps are computationally expensive, we proceeded to sort the results and filtered verbs that we judged likely to be irrelevant. Obviously, this is a highly subjective procedure and in a world of infinite resources we would have preferred not to do this filtering as it rested on our assumption about what is in the texts. Because the number of verbs in English were much smaller, we did not perform filtering for the St Croix data. After sifting, we were left with about 139.606 verb instances in Danish and an additional 6.459 in English.\footnote{Among the Danish verb instances are some duplicate, resulting from verbs being split into sub-word tokens.}

        In our workflow, these verb instances – linked to their particular place in a larger, cohesive textual unit – form our basic entity. They form the stuff we can subsequently make into something else.

    \section{Verb fermentation}
        The verbs are a starting point in the process of turning our sources into structured graph data relating to explicit actions. The core idea behind this operation stems from the GRAM-framework – Graph of Roles and Actions Model \cite{GRAM}. We do not employ the full framework but start from a crude skeletal version that might serve as a foundation for further graphing. A more extensive attempt is presented in the appendix. 

        In the GRAM-framework there are 3 types of nodes, with fundamentally different attributes.

        \begin{itemize}
			\item Actions. These typically have attributes relating to time of occurrence or duration. A given action is also tied directly to the part of the text that describes it. This links it to a verb instance.
            
            \item Actors. These include people but also objects and environmental forces as well as immaterial entities such as beliefs or stories. The type of actor is an attribute.

            \item Places. These might have attributes denoting coordinates of points or polygons, as well as some typology of geographical abstraction (village, city etc.).
		\end{itemize}

        \begin{figure}

            {\centering \includegraphics[width=0.45\textwidth]{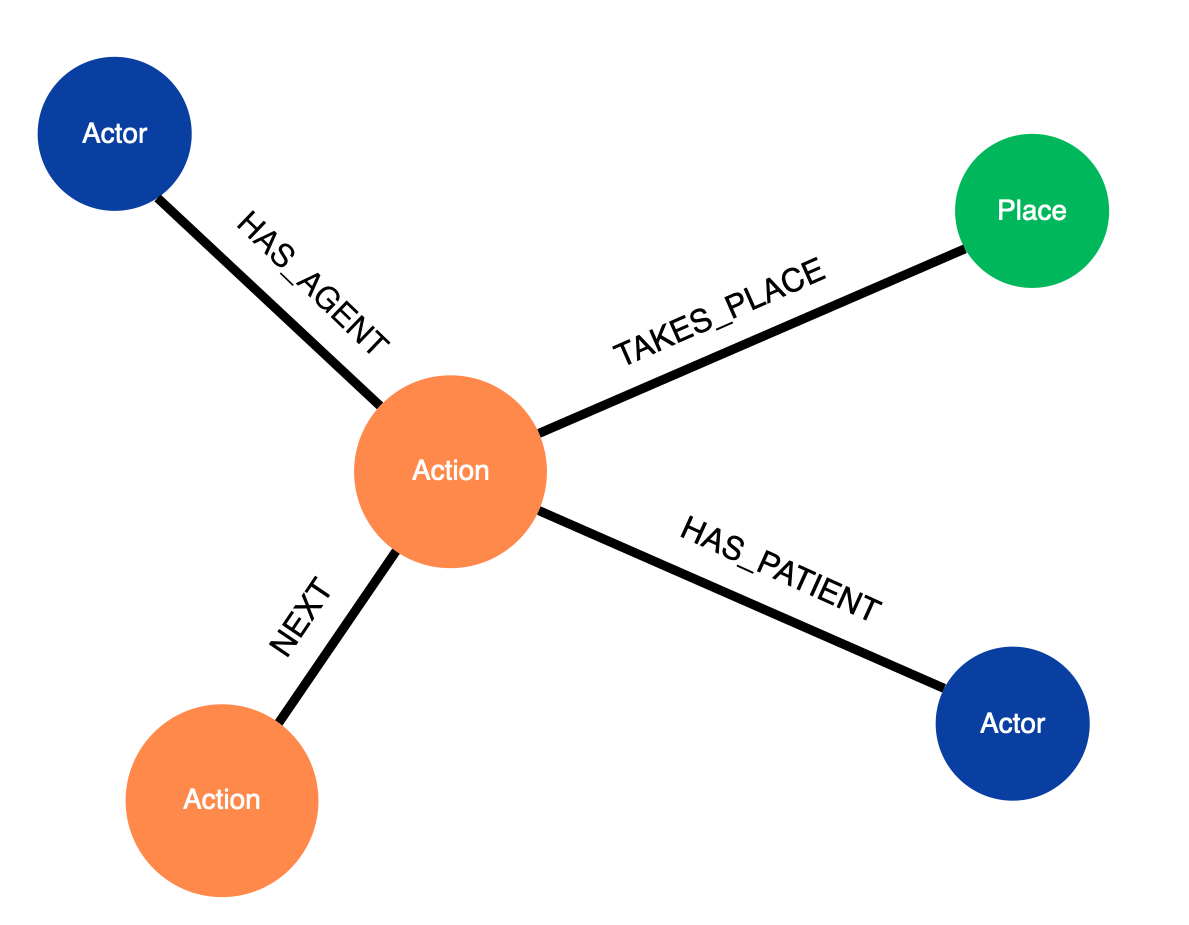}}

            \caption{Example of a simple GRAM-style graph.}

        \end{figure}  

        Nodes can be connected via different types of edges. For example, actions can be related to other actions with NEXT (denoting a temporal sequence described by the source) and to places with TAKES\_PLACE, HAS\_ORIGIN or HAS\_DESTINATION. Actors can be given roles in actions using HAS\_AGENT (when the actor is the one carrying out the action), HAS\_PATIENT (when they are the object of an action) or HAS\_INSTRUMENT (when the actor is a tool used to carry out the action). There are many other less commonly used edge types, but these are the ones from which we started our attempts at automated graphing. 

        Building GRAM-style graphs by hand is an exceptionally useful process. It allows you to get a clear sense of what is in a source as opposed to what you believe or infer to be in it. At scale, it also allows you to derive emic categories from corpora and to query based on such categorization. You get to know your sources exceptionally well. 
        
        The downside is that it is extremely labour intensive to apply to entire collections. Graphing the sentence of a deserter by hand might take a full day. And because building a graph involves a lot of human judgment on what is being said, outsourcing it to other people (student assistants for example) brings a lot of variation. Computers are less sophisticated, but they do not tire easily and they tend to be more systematic, even when wrong. For this reason our process consisted of manual graphing of select sources (guided by a lot of intuition) alongside an automated, simpler graphing done by a large language model. 

        If the previous steps could be framed with the metaphor of harvesting, threshing and milling, this step might perhaps be likened to the process of fermentation. It ultimately is a process in which something somewhat ephemeral is added (natural yeast, generative AI) that changes the properties of an otherwise static basis. Like fermentation, it is somewhat hard to control. If it goes well, it can be the basis of a wonderful thing (analysis, a sourdough loaf), but it can also turn sour fast. Sometimes you might not be able to pinpoint what exactly went wrong, but elements such as temperature and flour type are always important for the outcome.

        In order to have a workflow for prompting that allowed for consistent and structured output and easy changing of models (local and server-side), we opted to use the DSPy library in Python.\footnote{\url{https://github.com/stanfordnlp/dspy}} This library allows for structured input and output when running LLMs in loops. It also does minor adjustments of prompts to ensure consistency in output and optimization of performance.

        Our workflow consisted of two rounds of prompts, each processing every verb instance. The first prompt shows the model the full text. Inside the text, we wrap the verb to be analyzed in guillemets (which are never in the text to begin with), based on the extracted character spans from the part-of-speech-tagger. This is to help it locate entities, and is of obvious use if a text contains more than one instance of the same verb. For the prompt we describe a task and a role for the LLM as well as fairly minimal definitions of what we want it to find. We found that longer instructions or extended examples introduced erratic outputs and strange unintended biases. Even if the texts were all in archaic Danish, performance was better when prompting in English. 
        
        For the Danish texts, DSPy uses the following information to construct a prompt fitting of whatever model is specified:
        
        \begin{itemize}
        \item Role: You are an expert computational linguist specializing in 19th-century Danish text and semantic role labeling.
        \item Task: Extract action triples from runaway advertisement provided a NER-identified verb. You are presented with a text with one target verb occurrence wrapped in »guillemets« marking exactly which instance to analyze.
        \item Subject: The semantic subject of the marked verb — the specific entity performing THIS action.
        \item Title: The occupational title, judicial status or other emic label of the semantic subject of the marked verb.
        \item Verb: The full, reconstructed form of the marked verb.
        \item Object: The full semantic object of the marked verb specifically. If none output null.
        \item Infinitive: The infinitive form of the marked verb and ONLY the marked verb.
        \item Snippet: The full sentence containing the marked verb occurrence (without the » « markers in your output). Make no corrections to the original wording.
        \end{itemize}

        Unsurprisingly, the more ambiguous an element is, the harder it is for the model to retrieve. We iterated quite a lot to find a formulation in which the 'Title' – effectively the emic label assigned to an agent in the text – was identified. In many versions, the model struggled to include both legal titles (e.g. convict) and occupational titles (e.g. musqueteer). This confusion seems to be exacerbated by a fundamental ambiguity in many 'occupational' titles from the period, for instance 'peasant' which is in fact a legal category.
        
        We included both the full version of the verb, as well as the infinitive because we found it useful for later query.
        
        Perhaps the key element is, in fact, the extracted snippet. Here, we found that models acted very differently. Some included far too much text and some too little. We wanted a reasonable amount of text for there to be some context for the verb. It was not an explicit aim to have a snippet that contains all the information, for instance the name of the subject, as this would sometimes carry across too many sentences. We might think of the identified infinitive verb form as a label for that snippet, and for some approaches that combination alone might be enough to have the building blocks to do analysis, for instance via embedding and/or clustering.

        Models consistently struggled to identify the correct location of the action when presented with the full text which would often contain multiple locations. For this reason, we broke the extraction of location into a separate loop in which we presented the model with the snippet and the verb, and only presented the full text as context. This appears to have anchored the model's reading of the text.

       After testing a host of models, we chose to process the verbs with mistral-large-2512.\footnote{\url{https://docs.mistral.ai/models/model-cards/mistral-large-3-25-12}} It provided a reasonable balance between cost and performance and did so in both Danish and English.
       
        In the end, we get a tabular data structure in which each observation is a verb, with the following key variables: Subject, Title, Full verb, Infinitive form of verb, Object, Location, Snippet. Such a data structure comes ready for classification of all these separate variables. But more importantly, they form the basis for constructing small graphs of relations in which we learn something about what happened, who did it, to whom, when and where. This corresponds to the very basic graph structure outlined above.

        A logical next step would be to start disambiguating entities, to figure out if the servant doing x is also the servant doing y or who is the object z. However, as of the writing of this paper, we have found much use for this basic structure, which is relatively cheap and simple to deploy. Below, we will exemplify this by looking at a single type of action: to pretend. Our analysis relies on what has been extracted with the above-mentioned process. However, a loaf of bread is not a meal in itself, and as shown below, we incorporated manual annotation of the automated output into the analysis.

    \section{Acts of pretending}

        Pretending is often a key constituent of running away. After all, you need some sort of answer to the question 'who are you?' \cite{groebner}.
        
        In the material an act of pretending would often be described by the Danish verb 'foregive'. It does not strictly denote actions in which a person is lying about who they are, but it does suggest an element of distrust. The information given appears unverified.
        
        The process outlined above produces 836 observations of this particular verb. It appears in a range of forms, but all can be located because of the standardised infinitive. The data structure means that the subject is, in virtually all cases the agent, i.e. the person doing the pretending. Obviously, their titles vary, but often gives an initial impression of who they might be in the eyes of the author. Perhaps most importantly, the object-variable holds an entity that, in almost all instances, we can think of as an actor of the type 'story'. A common structure reads 'to be...' or 'to have been...'. 
        
        Initially, we planned to look only at this verb. However, upon embedding all snippets and reducing the vector space to fewer dimensions using umap, we got a clear indication that phrasings of 'foregive' were in many cases semantically similar to another verb 'udgive' (Fig. 3). Indeed, looking at actions described with this verb, we found a very similar formula with a story relating to identity being the object of the action. While this verb would not necessarily imply the same suspicion towards what a person claims to be, it often seems to do so given the context. In a few instances, 'udgive' also meant to circulate or to publish, which occurs in contexts of frauds and forgeries. We filtered these out manually, along with a few duplicate entries (a result of tokenization), and were left with a total of 1091 examples of pretending. Not all of them relate to actions performed during mobility – for example quite a few describe circumstances relating to strangers who have been caught in crime and whose identities appear in need of verification. However, all speak to ways of creating plausible narratives about identity, as well as ways of contesting such narratives.

        In this sense, the extracted snippet offer a way to compute pathways (understood as a distance metric relating to semantic similarity) within the topography of the new collection.

         \begin{figure*}

            {\centering \includegraphics[width=\textwidth]{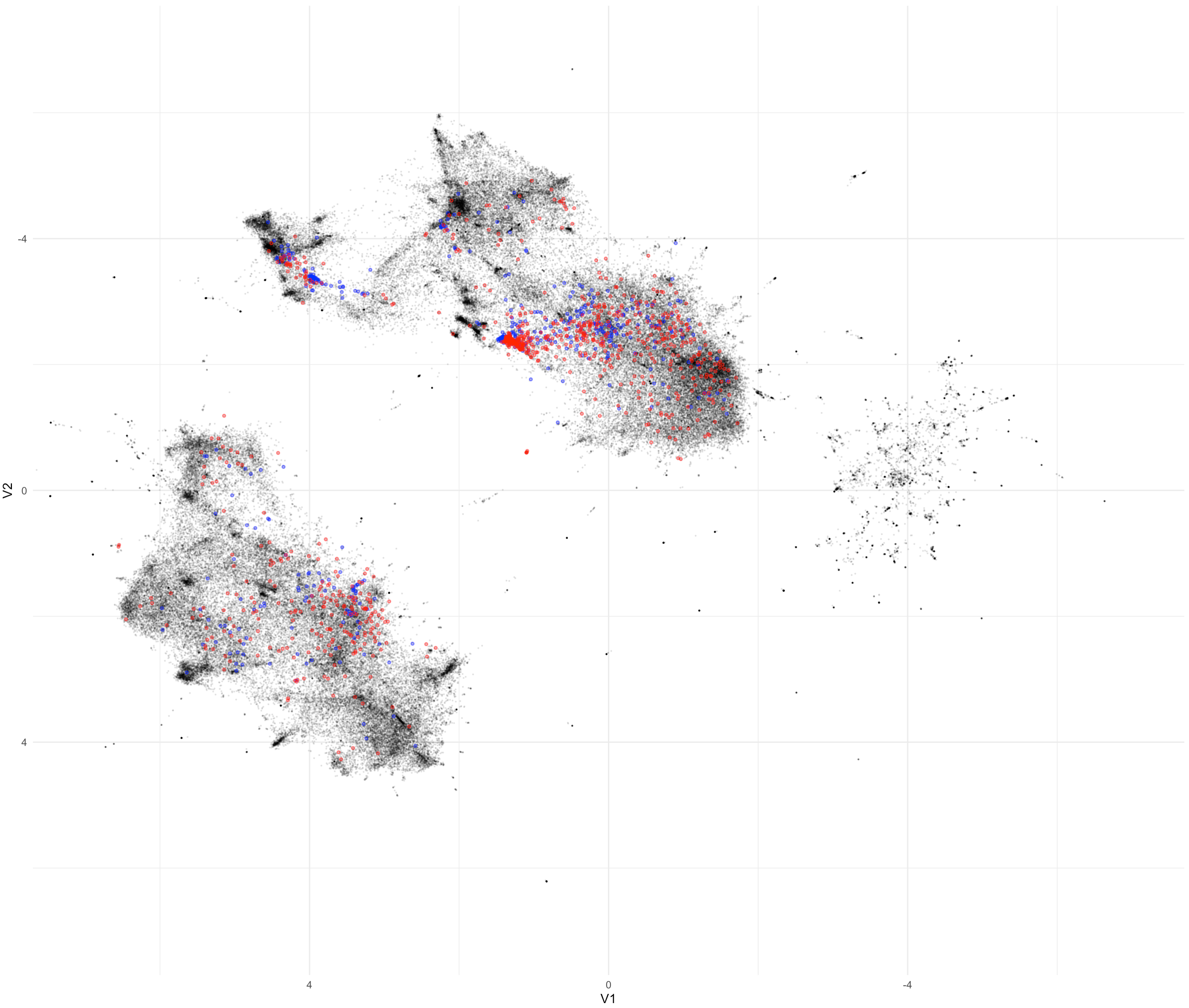}}

            \caption{Umap of all 139.606 Danish snippets embedded with '\textcolor{red}{foregive}' and '\textcolor{blue}{udgive}' highlighted. Embedding-model: DA\_SBERT\_Old\_News\_V1. Umap parameters: n\_neighbors = 25, n\_components = 3, metric = cosine.}

        \end{figure*}         

        In the advertisements from St Croix, we found only 3 instances of the verb 'to pretend'. Another eight instances were related via the verb 'to pass' as in 'to pass for x'. We included these 11 combined observations in the final dataset. We are not sure what to make of the relative scarcity of mentions of pretending in this material – if it represents an attribute of what running entailed in Denmark vs. the Caribbean, or perhaps if it relates to the way in which the colonial material narrates escape. In contrast, the colonial material is full of verbs relating to what the advertiser supposes or suspects the runaway to be doing. 

        Below are two examples both from the same 1870 police notice relating to a theft commited by a journeyman ropemaker in Northern Jutland.

        \textbf{Example 1:} Infinitive: To pretend (foregav); Subject: a journeyman stranger; Title: Journeyman; Object: that he was a journeyman ropemaker from Funen by the name of Mads Pedersen; Snippet: The person, that walked away with the goods, pretended to Vestergaard [the victim] that he was a ropemaker journeyman from Funen by the name of Mads Pedersen\footnote{Danish original text:\\Infinitive: Foregav; Subject: en fremmed Haandværkssvend; Title: Haandværkssvend; Object: at han var en Rebslagersvend fra Fyen ved Navn Mads Pedersen; Location: Fyen; Snippet: Personen, der bortgik med Sagerne, foregave til Vestergaard at han var en Rebslagersvend fra Fyen ved Navn Mads Pedersen}
        
        \textbf{Example 2:} Infinitive: To pretend (Udgive); Subject: he; Title: journeyman; Object: to be a mason; Location: Skjørbæk Inn; Snippet: whereas in Skjørbæk Inn he pretended to be a mason.\footnote{Danish original text:\\Infinitive: Udgive; Subject:  Han; Title: Haandværkssvend; Object: sig for Stenhugger; Location: Skjørbæk Kro; Snippet: hvorimod han i Skjørbæk Kro udgav sig for Stenhugger.}

        Generally, we see that in these instances, the LLM performed well, although there is a mistake in the first example, since the the location identified ('Fyen') is not the location of the action, but part of the story told. This is a recurrent mistake relating to these particular constructions in which the object is a story that often contains mention of locations. In other instances, we see the model get confused by occupational titles that form part of the story offered, but are contested by the context. Thus, sometimes the model accepts a title offered by the wanted person that the author of the text disputes.

        In order to get a sense of how subjects used different elements to craft plausible identities, we proceeded to manually code subjects and objects. For the subjects we annotated the gender of the pretender. For the objects we created variables relating to what the stories contain mention of. Here we noted when objects spoke of work or occupation, a geographic origin, a geographic destination, marital relations, physical injury/disease or misfortune (e.g. fire, shipwreck). We built these categories based on our cursory readings of what had been identified as the objects in the process. In this sense, the process aided the building of a coding scheme based on a sense of the data. Generally, most observations include at least one mention of something relating to the object variables, and in almost all cases the gender could be deduced from the subject variable. In this sense, the following coding, though manual, was much easier and faster than if it had involved reading entire texts filtered based on keywords. In cases of working with much larger datasets, it might be worthwhile to make an LLM do such coding, but we would have had no problem annotating much more data in this format.

        The 11 instances from the St Croix dataset did not fit the coding scheme well. Only one act of pretending involved an occupation (being a carpenter and a joiner) and only one mentioned an explicit origin (being from St Thomas). 2 instances, both relating to women, mentioned feigning physical injury (being 'lame'). Meanwhile, all except one revolve around being legally 'free', as opposed to enslaved. In the material from Denmark, freedom is only the explicit framing of the stories in a few instances, relating either to peasants during adscription (up until 1788) and convicts pretending to have been released. While the numbers relating to St Croix are small, the pattern is so striking that we feel somewhat confident in saying that pretending was a very different practice in a plantation colony than in Scandinavia.

        Gender also appears to factor heavily into how people pretended. Most notably, only about 20 percent of the instances when the pretender was a woman, do the stories related here revolve around work. This contrasts to 56.6 percent for men. Interestingly, marriage then appears to substitute for work, as marital status or relations are mentioned in 26.4 percent of observations with women as subjects, while in only 1.9 percent of the cases relating to men – in which case they are often talking about a woman who appears to have been present – effectively pretending on their behalf. Furthermore, the variables of work and marriage have a strong negative correlation, suggesting that you either say one or the other. Involving marital relations in the construction of an identity took a number of forms for these women. In some instances, it relates to where they come from, tying them to a household. Some related marriage by presenting themselves as widows. In other instances, marriage forms part of stories about destinations – women saying that they are on their way to a place because they are getting married. We also find instances where apprehended women present themselves as the wives of men in their company, though the context disputes their claims.

        Recent research has suggested that women formed integral part of the same labour markets as men in Northern Europe. The reality that women performed many of the same tasks as men, was a key conclusion of the work of Ågren et al, in their analysis of Sweden until 1800. \cite{agren_making_2017} We have no reason to suggest that Denmark looked different than Sweden in this regard, even if much of this data is from the nineteenth century. However, in creating identities tailored to the ears of suspicious audiences, women seemed to have played more on a gendered normative framework binding them to men, than on the realities of everyday work.

        In contrast, time seems to make only a negligible difference. Together, the ENO and SPOR datasets cover more than 130 years, during which society changed dramatically, including massive legal reforms changing the legal standings of commoners in relation to both the state and labour market. Yet, besides an increase in the frequency of mentions of work, the numbers are almost identical (Table 1). Thus, at a rudimentary level, the stories people presented about themselves when trying to pass for something they were not, contained similar elements across the period.

        \begin{table*}[]
            \centering
            \resizebox{1.95\columnwidth}{!}{%
            \begin{tabular}{|l|r|r|r|r||r|r|}
                 Theme & ENO (1750-1850) & SPOR (1867-1890) & Transcriptions (1720-1830) & St. Croix (1770-1810) & Men & Women \\ \Xhline{2pt}
                 Work (object) & 44.0 & 56.3 & 40.0 & 25.0 & 56.6 & 19.8 \\
                 Origin (subject) & 28.8 & 24.8 & 34.3 & 25.0 & 25.6 & 31.4 \\
                 Destination (subject) & 11.1 & 10.4 & 7.1 & 0 & 9.6 & 15.7 \\
                 Marriage (object) & 7.0 & 3.8 & 2.9 & 0 & 1.9 & 26.4 \\
                 Injury (object) & 2.9 & 2.2 & 1.4 & 50.0 & 2.4 & 3.3 \\
                 Misfortune (object) & 2.1 & 2.0 & 7.1 & 0 & 2.4 & 1.7 \\
                 Ownership (object) & 4.1 & 0.5 & 7.1 & 0 & 1.6 & 1.7 \\
            \end{tabular}
            }
            \caption{\footnotesize{Percentage distributions across the manually coded categories for each dataset. Corpuswide sex distributions to the right.}}
            \label{tab:placeholder}
        \end{table*}
        
        There was variation to the work identities deployed, however. We note an increase in the number of elite work identities employed in the later period. Liberalization of labour markets in the wake of democratization in 1849 and urbanization are likely to have played an influence, as work identities – especially those relating to skilled labour – became partially disentangled from judicial status. Yet, it is also hard to gauge if this shift in the data simply relates to the fact that the police gazette also contains more notices relating to fraudulent tricksters, and such con-artists often leveraged unusual and highly specific identities that they could somehow match credibly enough to scam someone. For instance, in the data is a man who in 1886 defrauded jewelers and watchmakers in Copenhagen by pretending to be history professor Kristian Erslev, \textit{the} founding father of historical source criticism in Denmark.

        Pretending could also take the guise of being victims of great misfortune. The pretending actions of 30 persons were explicitly tied to certain disruptive events that were out of their hands. Some would claim to be displaced Schleswigers, forced to move as a consequence of the events of the second Schleswig War of 1864. In 1886, 22 years after the war, that story was still viable, as an escaped poorhouse inmate wandered between jurisdictions pretending to be a Schleswiger with no home to return to, which seemingly prompted temporary poor relief wherever he went. The notion of being a 'deserving destitute', meaning to be without blame or responsibility in becoming poor, was also a recurring strategy. Another popular option within the theme of misfortune was pretending to be a shipwrecked sailor. Like the stories of displacement, this story could explain both your foreignness in any area you wandered, as well as your potential lack of papers and personal identification. However, pretenders likely had to be careful where they deployed this story, as residents living close to the site of the alleged shipwreck would know better. Similarly, in waters and straits with high degrees of oversight and control, like the Øresund Strait, this act of pretending would probably not do.

        To summarize, actions of pretending largely conformed to two main themes. One theme (and by far the most widely deployed) was to tap into structural societal expectations such as pretending ties to the labour market or marital ties to another person (of the opposite gender). When a man acts as a craftsman or a woman acts as a wife, pretending normality becomes a reflection of the society surrounding the pretending, and perhaps intends to evoke no reaction at all. This aligns with other studies of everyday tactics in the period \cite{Vilhelmsson_2020}. Another theme was pretending the extreme or the particular. These strategies were parts of elaborate, specific and exceptional backstories intended to trigger exceptional and specific responses – for example when a thieving prison workhouse dischargee from Lolland pretends to be a sailor making his way home to Zealand from a shipwreck in Gibraltar. While both themes were valid strategies, it certainly reads as though pretending normality was the easier task.

        In the short illustrative analysis above, we have approached the data armed with simple questions about the basic constituents of pretend identities. We have approached this by adding an additional layer of categorisation to allow us to think in terms of simple frequencies. However, we could just as well have used the framework as a straight-forward heuristics for qualitative analysis. Indeed, as we read through the 1102 instances, we found many narratives that prompted substantial follow-up questions. For instance, the objects identified frequently revolved around legitimation papers (passports, conduct books etc.), but often not always in the ways, we had anticipated. This was a society that required written documentation of your identity whenever you left your native parish, crossed key spatial thresholds or sought employment \cite{Krogh}. Thus, forgeries and misuses of papers were sometimes part of the crafting of identities. However, we also found instances when the strict demand for papers appears to have been used as leverage in pretending. For instance, in late November 1766 a servant named Else Kirstine Augustdatter left her employer not to return. She did so under the pretense that she wanted to go to Communion. For this, she needed her conduct book, which was held by her employer. She received it, but then vanished carrying a prop to be able to document her (modulated) identity, even if she had left service illegally.\footnote{\textit{Københavns Adresseavis}, 25 November 1766.} In other cases, we found examples of people who had left a place under the pretense that they would go back to somewhere to collect papers left behind. 
        
        Other stories speak similarly to the practice of exploiting expectations surrounding legal and illegal mobility. In at least two instances, we see runaways pretending to be military personnel searching highways for deserted soldiers. In one case, the deserters were pretending to be soldiers sent in search of their real selves. We knew this case already from having graphed it manually. In the other case, which we came across only through this automated workflow, we saw a twist on this ploy, as the identity of searching for deserters was even leveraged to defraud a person of a set of clothes, thereby inadvertently aiding the runner. In December 1806, a stranger in uniform had lodged in the advertiser's house under the surname Mikkelsen and had claimed to be an officer searching for a deserter near Aarhus. In the morning, he had demanded a set of clothes in order to appear 'unrecognizable for the deserter, thereby he would catch him sooner'. He pawned his uniform for a set of civilian clothes. Three weeks later, he had not returned and was advertised. It appears likely that he was also himself the deserter, and by changing clothes, his chances of further success would have risen.\footnote{\textit{Aarhuus Stifts Adresse-Contoirs Tidender}, 27 December 1806.}

	\section{Conclusion}
        After having been harvested, threshed, milled, and fermented, our sources suggest a living, unruly culture of improvisation: Deserters pretending to be on the hunt for deserters, women tying themselves to husbands who may never have existed, a servant slipping out the door with her conduct book and a story ready. These are not patterns that a keyword search alone would have surfaced, nor ones that unaided close reading across two centuries of sources could easily have assembled. 
        
        We believe that this pipeline and its future developments provides a robust starting ground for an action-oriented and granular social history. Importantly, it does so in ways that blur the distinction between quantitative and qualitative analysis. Auto-GRAMs lend themselves to both emic and etic categorisation while also serving as a heuristics to find materials for a prolonged engagement with those parts of a corpus that conjure up new questions.

        Most importantly - as we have attempted to make clear during the article - any adaptation of the harvest-thresh-mill-ferment workflow must take into account that it is by no means a linear or copy-pasteable process. As with literal processing of grains, the outcome of each step is affected by both the makeup of the grains themselves and the specific techniques applied to them. Different grains have different protein and starch contents, and milling the grains differently can accentuate specific characteristics of their composition. This, in turn, affects the structure of the dough and the outcome of the fermentation, which also introduces new variables such as humidity and temperature. By this metaphor, we mean to say that everything in the described workflow is connected and interdependent and must be developed as such. Each step inherits the product of its prior step, which then introduces new elements to deal with. As sources and contexts differ, each processing step must adapt.

        Our choice of a guiding metaphor in this working paper is not just a matter of easing the imagination with a playful gesture. Digital history has often come steeped in extractive metaphors ('mining' and 'cleaning' is ubiquitous), bringing along an air of positivism. We have attempted to stress, that this is neither a neutral or purely subtractive process. Firstly, each step radically transforms the material bringing about loss, but also added value that cannot be thought of only in terms of purification. Further, the addition of language models (both BERT-based and generative), means to rely on already trained weights that aid, but also affect the textures of our research data. We have attempted to highlight these themes of both subtractive and additive transformation with a different set of metaphors that might helps us think about what happens to the substrate of historical sources as they become the stuff that might feed the writing of history.

    \section{Acknowledgments}
	
		We thank Matias Kokholm Appel and Frederik Larsen for valuable technical sparring, and Anders Dyrborg Birkemose and Kamilla Matthiassen for inputs in the process.

    \newpage
    \bibliographystyle{abbrv}
    \bibliography{bibliography}

    \newpage

    \section*{Appendix: A more extensive automated graphing}

        A social historian might want a more elaborate graph structure than what we presented above. Indeed, the original GRAM-framework presents such an elaboration. Originally designed as a tool-set for manual processing of sources, LLMs can get us some of the way there. This appendix briefly details our own attempt along this path.
        
        We used the St Croix dataset as the basis for a series of prompts that takes the two-step prompt approach outlined above as its basis, but extends it further.\footnote{All code available at \url{https://github.com/HisLabAAU/autoGRAM/}} To the initial prompts (1. Subject-object-snippet; 2. Location), we added the following sequence:

        \begin{itemize}
        
			\item A prompt that asks if the snippet mentions any tools/instruments as part of the action. As with the locations, doing this as a separate step anchored in the snippet appears to reign in the LLM-response. We automatically relate these findings to the action with the HAS\_INSTRUMENT relation type.
			
			\item A step designed to disambiguate the identified subjects. This step is key, as it allows us to see the same entity relating to multiple actions and, eventually, with multiple roles across a text. The prompt was structured so that the LLM assigns an id to what it perceives to be each unique subject in the previously derived subject variable. It sees the full text as context for this disambiguation task. It also assigns a 'canonical' label to the entity, typically the name under which an entity is most often represented.

            \item A step similar to the one above, but targeting the object. Entities found in the object variable are matched against the disambiguated subject-entities. If there is a match, the assigned id and label carry over, but if not, a new id and label are assigned.

            \item Role labeling. Based on the entities disambiguated in the previous steps, each entity is assigned a role from the GRAM-vocabulary in relation to every action in which they figure as either subject or object.\footnote{For the prompt we framed the roles like this: HAS\_ADDRESSEE: the entity receives a message through the action and is thereby its audience; HAS\_AGENT: the entity is the agent performing the action itself; HAS\_PATIENT: the entity is the patient of the action (often the semantic object); HAS\_BENEFACTIVE: the entity derives a benefit from the action but is not its agent; HAS\_COMITATIVE: the entity is performing the action together with others, becoming co-agents of the action; HAS\_EFFECTOR: the entity triggers or causes the action but is not the agent (e.g. the person who orders the action, or an environmental event that causes the action); HAS\_EXPERIENCER: the entity observes, perceives, or senses the action, but is neither its agent or its addressee (e.g. a feeling); HAS\_INSTRUMENT: the entity is a tool used in the action; HAS\_MEDIATOR: the entity serves actively as an intermediary between the agent and the patient; HAS\_MEDIUM: the entity is the material means through which mediation is realized (e.g. a contract, an amount of money, a transportation); HAS\_RECIPIENT: the entity receives something tangible through the action (as possessor); HAS\_RESULT: the entity comes into existence through the action; HAS\_TARGET: the entity is the yet unrealized target or ambition of action; HAS\_THEME: the entity is the topic of verbal action (e.g. of a discussion or a narrative).}

            \item Role labeling of locations. For the places, we did a similar prompt, but with place labels.\footnote{Framed in the prompt like this: TAKES\_PLACE: the location is where the action itself takes place; HAS\_DESTINATION: the location is mentioned as the destination of an action, typically involving mobility; HAS\_ORIGIN: the location is the spatial starting-point of an action that has a destination (that might or might not be mentioned); IS\_SUBJECT\_OF: the location is mentioned as the location of a past, future or hypothetical action. Note that the IS\_SUBJECT\_OF label is our in-house addition to the GRAM-vocabulary.}
	
		\end{itemize}

        \begin{figure*}

            {\centering \includegraphics[width=1\textwidth]{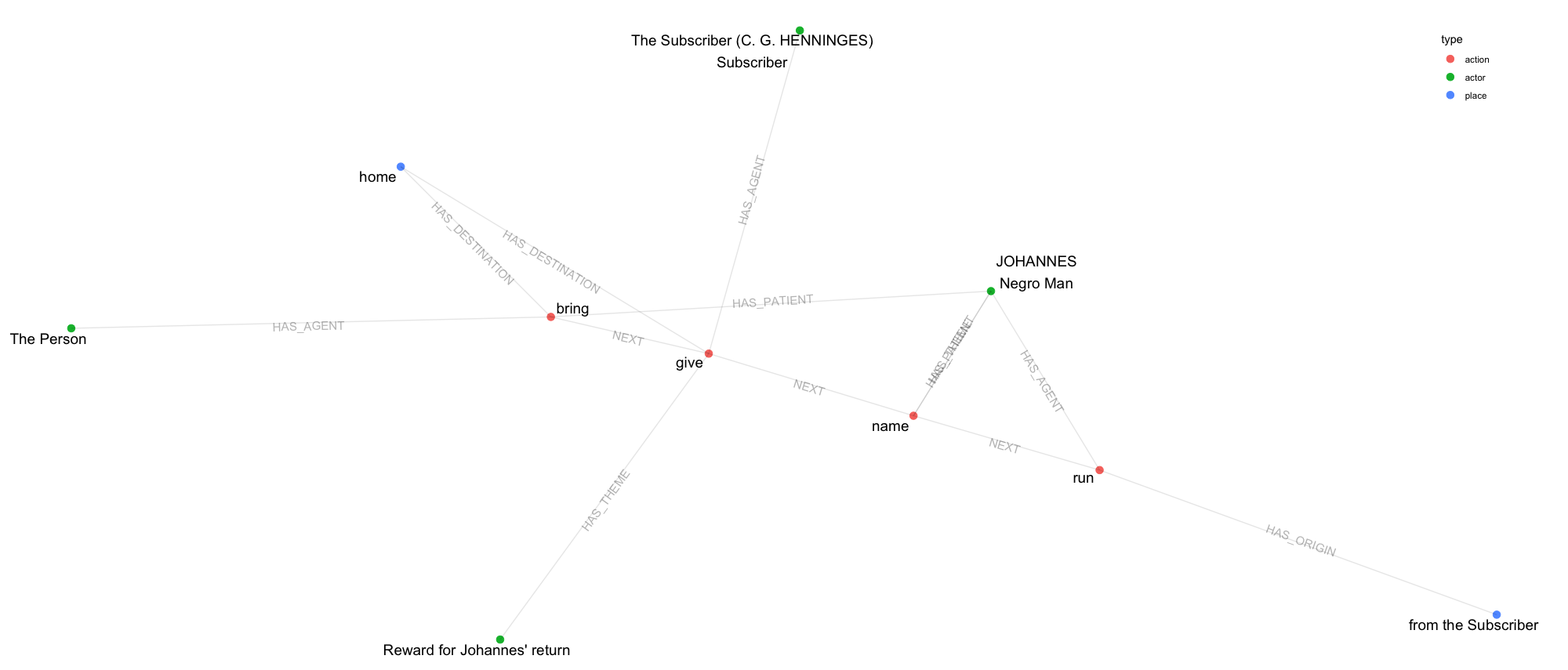}}

            \caption{Example 1.}

        \end{figure*}

        The resulting data structure is tabular with all of these elements split across columns. For the sake of simplicity, there is no disambiguation of instruments and locations. A simple script wrangles the data into graph form, with nodes and edges separated into discrete tables. On a few occasions, illustrated in the examples below, an entity is twice related to the same action, which happens if they appear to be both its subject and object. We initially tried to also prompt the model to put actions in their perceived chronological order, but it did not work well enough to be worthwhile. Instead, we employed the NEXT relation type mechanically, relating simply to an action's order of appearance in the text.

        Let us look at some examples, by processing three randomly pulled texts from the dataset.

        \subsection{Example 1}        

            \begin{quote}
            \textbf{25 July 1805:} 
            Ran away from the Subscriber a few Days ago, His Negro Man named JOHANNES, formerly the property of Mr. HANS BRETFELDT. The Subscriber will give a Reward to the Person who will bring him home. C. G. HENNINGES. Christianstæd, 25th July 1805.
            \end{quote}

        \begin{figure*}

            {\centering \includegraphics[width=1\textwidth]{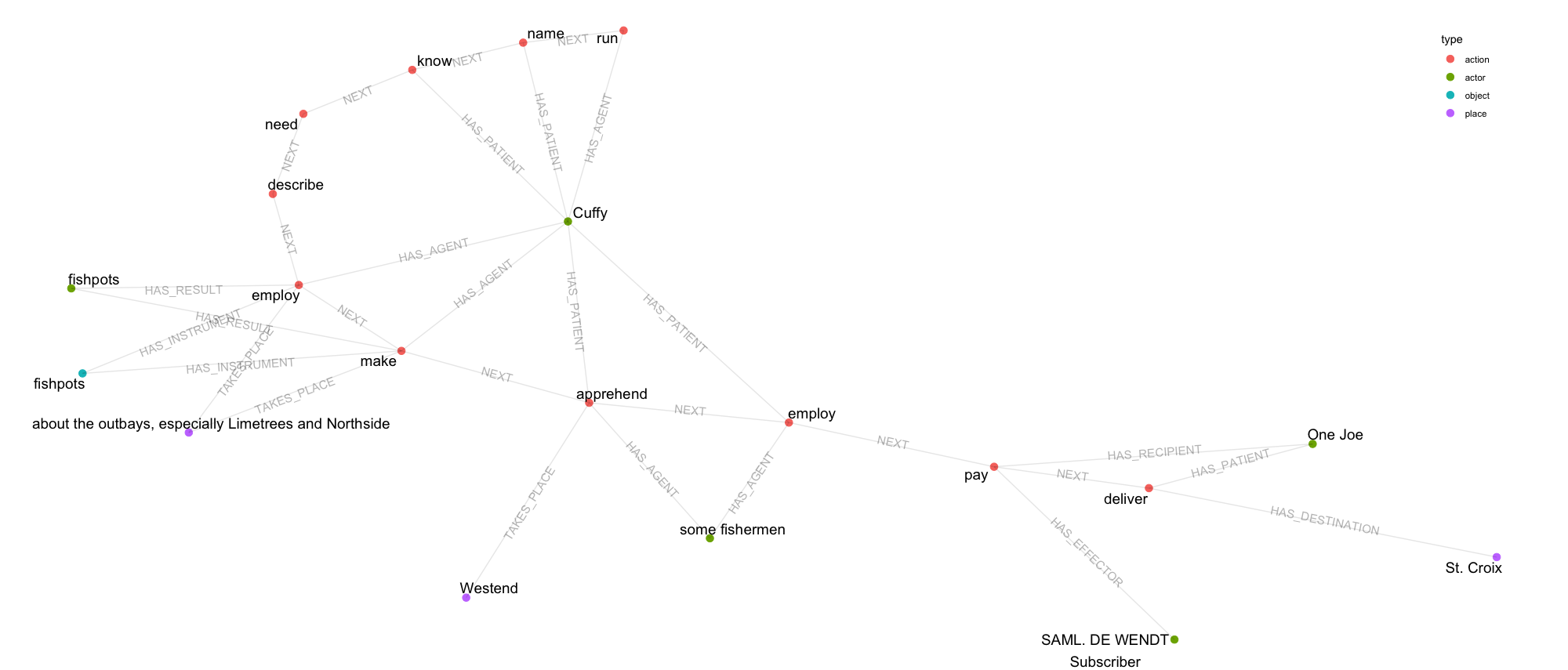}}

            \caption{Example 2.}

        \end{figure*}

        This is a very simple advertisement and inspecting the graph (figure 4) the model appears to have done almost entirely what we hoped for it to do. It has disambiguated entities well, mostly understanding the roles of Johannes (whose node label consists of the canonical label assigned by the LLM then, below, the emic label). It is debatable if it makes sense for 'from the Subscriber' to be understood as a location, but since we aim for the location extraction to be inclusive, this is the behaviour we wanted. Similarly, the place 'home' is given the role of 'destination' for the verb 'give' relating to the reward, but it would have made better sense if this had been labeled as IS\_SUBJECT\_OF given the future, hypothetical nature of that payment. One might argue that Johannes is perhaps the 'theme' of the action 'to give', though that takes some interpretation, and the prompt is mostly designed to reign in interpretation due to the risk of hallucination. We might have wished for the model to also make 'The Person' (a hypothetical entity apprehending Johannes) the recipient of the reward associated with the action 'give'. Johannes is both 'theme' and 'patient' of the action 'name', but only the latter makes sense. However, this graph mostly reflects the text and does so in very direct way. 

         \vspace{1cm}

        \subsection{Example 2}

            \begin{quote}
            \textbf{29 July 1803:} 
            Run away from the Subscriber, A Negro man named Cuffy. He is notoriously known, therefore need to description. He is generally employed about the outbays, making fishpots, especially Limetrees and Northside: he was lately apprehended at Westend, and employed by some fishermen. One Joe reward will be paid by delivering him to SAML. DE WENDT. St. Croix, 26th July 1803.
            \end{quote}

            \begin{figure*}

            {\centering \includegraphics[width=1\textwidth]{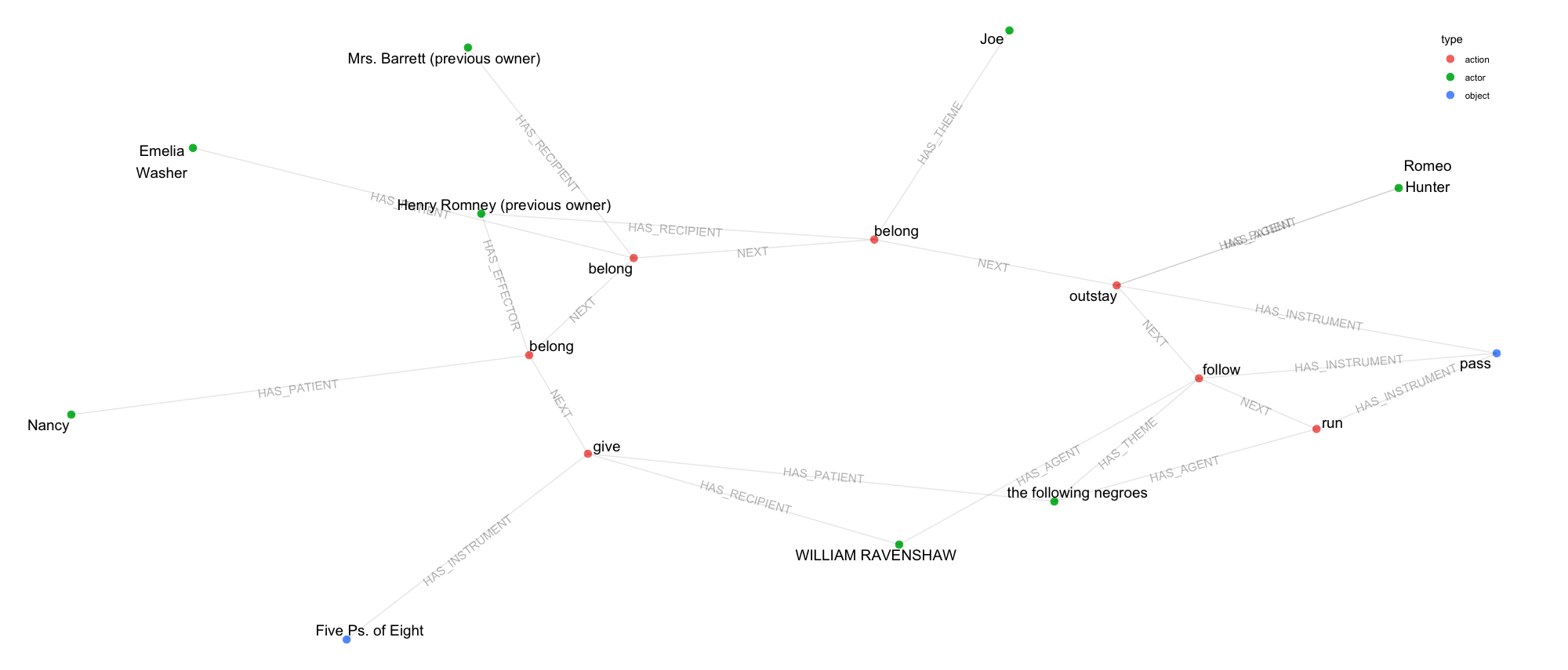}}

            \caption{Example 3.}

        \end{figure*}

        Looking at the results (figure 5) from this somewhat more complicated text, we see that the model again did a good job at extracting and disambiguating entities. It also understands that Cuffy is not only agent of many of the actions, but also patient of actions such as 'to apprehend' and 'to employ'. The 'hallucinated' verb 'describe' is not in the text but owes to a misunderstanding on the part of the part-of-speech tagger. The LLM correctly identifies the fishpots mentioned as part of his work as a result of the action and locates them correctly. They are doubled because they have been identified both in the instruments and objects passes, an error that would be solved by an additional step disambiguating instruments, in a similar way to objects. The reward of 'One Joe' (a standard monetary amount) is labeled as 'recipient' of 'pay' and 'patient' of 'deliver', both which are somewhat convoluted ways of thinking the relations. Still, the graph is meaningful.

        \subsection{Example 3}

            \begin{quote}
            \textbf{5 August 1776:} 
            RUN AWAY from the subscriber, the following negroes, viz. Romeo, who has a pass as a Hunter, but has out-stayed his time; Othello, a Cooper; Joe, who did belong to Henry Romney; Emelia, a noted Washer, who did belong to Mrs. Barrett, and Nancy, who also did belong to Henry Romney. Five Ps. of Eight for each negro will be given on delivery to. WILLIAM RAVENSHAW.
            \end{quote}

        This text is much more complicated (figure 6). Aside from the incorrectly identified verb ('follow' derived by the pos-tagger from 'following'), the text also contains an element that the model has not been instructed on how to handle: grouped entities, in this case the grouped mention of five runaways ('the following negroes'). For it to make sense, the individual runaways would each have to be linked to the group\footnote{In the GRAM-framework this would use the IS\_PART\_OF label.} or they should each have been made co-agents of the action 'run'. However, both of these solutions would imply that they ran together, which the text is not entirely clear on. The model, however, infers this and even makes the pass carried by Romeo an instrument for the group itself. It also, correctly, makes the pass the instrument of his particular action 'outstay'. The group is, however, correctly associated wit the verb 'to give' (a reward), though the role ('patient') is misleading. William Ravenshaw's role is also partially misunderstood, as he is the agent not recipient of the reward-giving. The roles of Mrs. Barrett and Henry Romney are also obscured ('recipient' of Emelia and 'Joe', with Romney also being 'effector' of Nancy belonging to him). The LLM has deduced that Romney and Barrett are 'previous owner', though this is not how we understand the text.
        
        Clearly, this text, and the dense passage about belonging, makes role labeling harder for the model. Most links make sense, but how the model interprets them would be a problem to querying based on action roles. Most problematic, however, is the fact that Othello is entirely absent from the graph. The reason is obvious – he is not directly tied to a verb. This reveals a pitfall in the heavy reliance on verbs to untangle the texts.

        Overall, though example 3 is shakier than the others, all three graphs convey meaning from the texts in a structured way. It helps, somewhat, that the texts are in English, as LLMs are notoriously better at high-resource languages than others. A cutting-edge model would likely have performed even better. Role disambiguation is obviously difficult, especially beyond 'agent' and 'patient' and tweaking the prompts to include elements such as place disambiguation or group association might bring further precision.
        
        Graphs like these allow broad ranging querying: for instance, what are the instruments relating to a given type of action? Who are agents and who are the patients in a text corpus? Under what circumstances do these roles flip? And where do actions with specific combinations of actors and roles happen? While perhaps not reliable enough to form the basis of robust generalization, automated graphs such as these would help research as a heuristics. Their uses are specifically wide-ranging for a granular social history interested in what social relations look like in terms of situated practices. However, it is, perhaps simultaneously, limited by the fact that it does not name or delimit the social relation. This relates to the approach's dependency on the explicitness of the sources and the fact that, in historical sources broadly, abstractions are rarely figured as actors in and off themselves. Indeed, in the examples given here, enslavement, ownership and asymmetry are all expressed only indirectly through verbs ('belong', 'employ', 'apprehend', perhaps 'name') and emic labels (here racial). As a tool for granularity, the framework seems to suggest that the abstractions that guide social interactions are real through a range of specific configurations of roles, not themselves as entities.
	
\end{document}